\documentclass{aa}
\def\et{et al.\ }                
\def\1928+738{QSO\,1928+738}    
\def\2007+777{BL\,2007+777}    
\begin{document}

\thesaurus{11(05.01.1; 03.20.2; 11.17.4 \1928+738; 11.02.2 \2007+777)}

\title {VLBI differential astrometry at 43\,GHz}

\author{J.C.~Guirado\inst{1}
\and J.M.~Marcaide\inst{1}
\and M.A.~P\'erez-Torres\inst{1} 
\and E.~Ros\inst{2} }

\offprints{J.C.~Guirado}

\institute{
  Departamento de Astronom\'{\i}a, Universitat de
Val\`{e}ncia, E-46100 Burjassot, Valencia, Spain
\and
  Max-Planck-Institut f\"ur Radioastronomie, Auf dem H\"ugel 69, D-53121, Bonn, Germany }

\date{A\&A, 353, L37-L40 (2000) } 
\maketitle

\begin{abstract}

From 43\,GHz VLBA observations of the pair of radio sources \1928+738 and 
\2007+777 we have demonstrated the feasibility of precision phase-delay differential 
astrometric techniques at millimeter wavelengths. 
For a pair of sources with 5\degr\, separation and high 
antenna elevations, we have shown that present astrometric models and millimeter 
arrays are advanced enough
to model the differential phase-delay to within 2 picoseconds, less than one tenth 
of a phase-cycle at 43\,GHz. The root-mean-square of the 
differential phase-delay residuals is dominated by the 
fluctuations of the atmospheric water vapor. 
We have determined the relative position of the observed sources with a precision 
twofold better than previous determinations at lower frequencies and, more 
importantly, largely free from ambiguous definitions of the reference point on the structure of 
the radio sources. Our result makes 43\,GHz VLBI phase-delay differential astrometry an ideal tool 
to study the absolute kinematics of the highly variable structures of 
regions near the core of extragalactic radio sources. 


\end{abstract}

\markboth{J.C. Guirado et al: VLBI Differential Astrometry at 43\,GHz}{}

\section{Introduction}

One of the trends in Very-Long-Baseline Interferometry (VLBI) is to augment the angular
resolution of the observations in search for a more detailed 
view of the inner structure of extragalactic radio sources.
This is effectively carried out by either observing at millimeter wavelengths 
(mm-VLBI) or, at cm-wavelengths, by combining ground telescopes with antennas 
in space. The correct interpretation of these high-resolution observations is 
of much relevance
since they map the morphology of highly variable regions close to the 
central engine of AGNs. However, multi-epoch analyses directed 
to understand the dynamical behavior of these inner 
regions critically depend on the alignment of the images: no solid  
conclusions can be extracted without an accurate source component (i.e. core) 
identification. In particular, VLBI reveals that cm-wavelength components 
break up in complex structures with multiple features at mm-wavelengths. 
These compact mm-features 
show a strong variability, which may be the result of phenomena only seen so far in 
numerical simulations (G\'omez et al. 1995). 
For a meaningful physical understanding of those compact features, a detailed 
knowledge of the (absolute) kinematics 
of the region is crucial. It is therefore highly desirable to extend 
precision differential phase-delay 
astrometry to mm-wavelengths.\\
In this Letter we demonstrate the feasibility of using phase-delay differential 
astrometry at 43\,GHz. We have selected the pair of sources \1928+738 and 
\2007+777, $\sim$5\degr\, apart, with flat spectra, high flux densities, and 
rich structures at 43\,GHz. The astrometry analysis of these data show 
the advantages and possibilities of mm-wavelength differential astrometry. 

\section{Observations and Maps}

\begin{figure}[t]
\vspace*{313pt}
\includegraphics{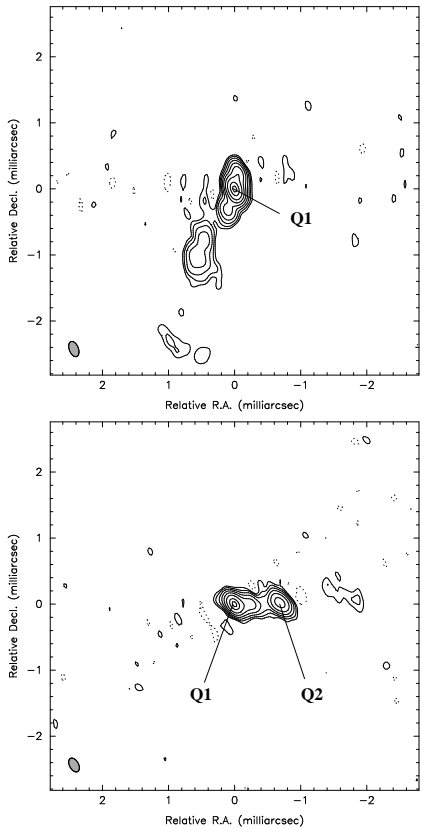}
\caption[]{43\,GHz hybrid maps of \1928+738 (left) and \2007+777 (right) at
epoch 1999.01. Contours are -0.5,0.5,1,2,4,8,16,32,64, and 90\% of
the peak of brightness, 0.56 Jy/beam, for \1928+738, and
-1,1,2,4,8,16,32,64, and 90\% of
the peak of brightness, 0.22 Jy/beam, for \2007+777.
The restoring beam (shown
at the bottom left corner of each map) is an elliptical Gaussian of
0.24$\times$0.14\,mas (P.A. 21\degr) for \1928+738, and
0.23$\times$0.14\,mas (P.A. 29\degr) for \2007+777.
The astrometric reference points are labeled as
Q1, and correspond to the peak of brightness of each map. See text.}
\end{figure}

We observed the radio sources \1928+738 and \2007+777 at 43\,GHz on 1999 
January 3 from 15:00 to 23:30 UT. We used the complete Very Long Baseline 
Array (VLBA) recording in mode 256-8-2 in left circular polarization to 
achieve a recording bandwidth of 64\,MHz. 
We interleaved 
observations of \1928+738 and \2007+777 using integration times between 
40 and 130 seconds on each source to make total cycle time durations 
between 110 and 300 seconds (antenna slew time was 15 seconds). 
The data were correlated at the National Radio Astronomy Observatory 
(NRAO, Socorro, NM, USA). Detections were found on both sources to all 
stations but Hancock, presumably due to severe weather conditions at this site. 
We made manual phase calibration, visibility amplitude
calibration (using system temperatures and gain curves from each antenna),
and fringe fitting at
the correlation position with the NRAO Astronomical Image Processing
System (AIPS). 
For astrometric purposes, we further processed the data in AIPS
(tasks {\sc mbdly} and {\sc cl2hf}) to obtain, for 
each baseline and epoch, estimates of the group delay, phase-delay rate, and 
fringe phase at a reference frequency of 43,185\,MHz. 
We discarded data from Saint Croix and
Pie Town stations, which showed unacceptable scatter in the observables, for the 
astrometric analysis presented in Sect. 3.
For mapping purposes, we transferred the data into the Caltech imaging 
program {\sc difmap} (Shepherd \et 1995). We performed several iterations of 
self-calibration in phase and gain. We present the resulting hybrid maps in Fig. 1.\\
The 43\,GHz map of \1928+738 displays several  
jet components extending southwards. All these features appear blended together 
as only one or two components in previous maps at cm-wavelengths (Guirado \et 1995, 
hereafter G95; Ros \et 1999, hereafter R99), and those obtained with space VLBI 
(Murphy \et 1999).  
The brightest knot, labeled as Q1, is probably a jet component, unless the 
source is two-sided, but it is compact and well defined. Thus, it constitutes
an appropriate reference point for relative astrometry at a 
single epoch. However, this component would not be a suitable reference point 
for a multi-epoch comparison of the relative separation between the two 
sources as it is likely to move and evolve in brightness and shape 
over time.\\ 
The 43\,GHz map of \2007+777 represents a significant improvement in the 
knowledge of the inner structure 
of this source (see Fig. 1). The brightest knot seen earlier at cm-wavelengths 
(Guirado \et 1998, hereafter G98; R99) breaks-up in at least three new features. 
The kinematic nature of component Q2, almost 
as bright as the easternmost component Q1,
is of much interest. The brightest feature of \2007+777 at cm-wavelengths, 
a blending of all components seen within 1\,mas from the origin in our 43\,GHz map,
has been taken as a reference point for astrometry (G95; G98; R99); even more, 
this feature has been considered stationary for 
multi-epoch astrometry analyses. Accordingly, should component Q2 be a 
travelling knot, the selected reference point at cm-wavelengths is likely
to be not stable over time and part of previous astrometry results must 
be revised. 

\section{Astrometry Analysis}

A goal of this research has been to calibrate the limitations of our standard astrometry 
procedure for 43\,GHz, as well as to study the potential precision of the astrometric data 
at this frequency. Therefore, the data-reduction procedure for the 43\,GHz observation 
deliberately followed the same steps as those used for the 5\,GHz (G95) or 
8.4\,GHz observations (G98; R99). We briefly go again over  
each step of this analysis: For our 43\,GHz data, (i) we predicted,
via a precise theoretical model of the geometry of the array and the propagation 
medium, the number of cycles of phase between
consecutive observations of the same source to permit us to ``connect" the phase delay
(e.g. Shapiro \et 1979; G95; R99);
(ii) we defined as reference points in the 43\,GHz images of the two radio sources
the maximum of the brightness distribution (components Q1 in the maps of \1928+738 and 
\2007+777; see Fig. 1) and subtract the contribution of the structure of the 
radiosources, with respect to the reference points selected, from the phase 
delays; (iii) we formed the differenced
phase delays by subtracting the residual (observed minus theoretical values)
phase delay of \2007+777 from the previous observation of \1928+738; and
(iv) we estimated the relative position of the reference points 
from a weighted-least-squares analysis of the differenced residual
phase delays. For this analysis we used an improved version of the 
program VLBI3 (Robertson 1975).

\noindent
In step (i), the geometry of our theoretical model (set of antenna coordinates, 
Earth-orientation parameters, and source coordinates) was consistently taken 
from IERS (IERS 1998 Annual Report, 1999). The theoretical model also accounted for the 
effect of the propagation medium in the astrometric observables. We modeled the 
ionospheric delay by using total electron contect (TEC) data from GPS-based global ionospheric maps 
generated at the epoch of our observations by the Center for Orbit
Determination in Europe (Schaer \et 1998). We followed the geometric corrections described 
in Klobuchar (1975) and Ros \et (2000). 
We modeled the tropospheric zenith delay at each station as a
piecewise-linear function characterized by values specified at epochs
one hour apart. We calculated a priori values at these nodes from
local surface temperature, pressure, and humidity, based on the model
of Saastamoinen (1973). The antenna elevations were always higher than 
20\degr\, at all stations; this allowed us to use the dry
and wet Chao mapping functions (Chao 1974) to determine the
tropospheric delay at non-zenith elevations for each observation at each site. 
We estimated the tropospheric zenith delay at the nodes of each 
station, along with the relative position of the sources, 
from a weighted-least-squares analysis. 

\section{Results and Discussion}

\begin{table}[t]
\caption[]{Contributions to the standard errors of the
estimates of the coordinates of \1928+738 minus those of \2007+777
( $\delta\Delta\alpha$, $\delta\Delta\delta$) from the
sensitivity study. }
\begin{flushleft}
\begin{footnotesize}
\begin{tabular}{llccc}
\hline
  Effect  & & Standard      & $\delta\Delta\alpha^b$    & $\delta\Delta\delta$ \\
          & & Deviation$^a$ &     ($\mu$s)  &  ($\mu$as)   \\
\hline
Statistical errors$^c$    &             & --               &  4   & 25 \\
Ref. point identification &             & --               &  0.6 &  1 \\
Station coordinates       &             &  2\,cm           &  6   & 25 \\
Coordinates               & $\alpha$    & 100\,$\mu$s      & 23   & 41 \\
of \2007+777:             & $\delta$    & 300\,$\mu$as     &  9   &  9 \\
Earth's pole:             & $x$         & 150\,$\mu$as     &  2   &  9 \\
                          & $y$         & 250\,$\mu$as     &  2   &  4 \\
UT1-UTC                   &             &  15\,$\mu$s      &  3   &  5 \\
Earth's nutation:         & $d\psi$     & 170\,$\mu$as     &  0.3 &  1 \\
                          & $d\epsilon$ &  80\,$\mu$as     &  1   &  2 \\
Ionosphere$^d$            &             & 1 TECU           &  1   &  2 \\
\hline
rss$^e$                   &             &                  & 26   & 56 \\
\hline
\end{tabular}
\end{footnotesize}

\begin{scriptsize}
\noindent
$^a$  The standard deviation of the fixed geometrical parameters of our astrometric model 
(all entries but ionosphere) 
were taken from IERS Annual Report 1998 (1999). The 2\,cm standard deviation of the site coordinates 
corresponds to each of the three coordinates for each antenna site.\\ 
$^b$  Notice that the values of $\delta\Delta\alpha$ are in $\mu$s. To convert $\mu$s to $\mu$as, the factor 
15$\cdot\cos\delta_{\rm{QSO}\,1928+738}$ ($\sim$4.2 ) must be used.\\
$^c$ The statistical errors include the uncertainties of the tropospheric zenith delays at the nodes of 
the piecewise linear function used in the troposphere model (see Sect. 2).\\ 
$^d$ Standard deviation provided by the global ionospheric maps at each site. 1 TECU = 1$\times$10$^{16}$ el\,m$^{-2}$.\\
$^e$ Root-sum-square of the tabulated values. 
\end{scriptsize}
\end{flushleft}
\end{table}

\noindent
From the astrometric analysis described in Sect. 3, we obtain the following 
J2000.0 coordinates of \1928+738 minus those of \2007+777 at 43\,GHz:\\

\noindent
\begin{tabular}{lll}
$\Delta\alpha=$ & $-0^{h}\,37^{m}\,42\rlap{.}^{s}503443$ & \, $\pm\,0\rlap{.}^{s}000026$\\ 
$\Delta\delta=$ & $-3^{\circ}\,54'\,41\rlap{.}''677208$   & \,  $\pm\,0\rlap{.}''000056$\\
\end{tabular}

\vspace*{0.2cm}
\noindent
where the quoted uncertainties are overall standard errors (see Table 1), nearly twofold 
smaller than the standard errors corresponding to previous determinations at 
5 and 8.4\,GHz. 
From the comparison of the results of the sensitivity analysis 
displayed in Table 1 with similar sensitivity analyses at lower
frequencies (G95; R99), 
we see that the improvement in precision comes from
(i) the small contribution to the standard errors 
of the reference point identification in the map (dominated by image noise), as 
a consequence of the improvement of resolution of the maps and of the 
lack of ambiguity in selecting the components acting as reference, and 
(ii) the negligible contribution of the ionosphere, that 
scales down by a factor of 25 with respect to its contribution at 8.4\,GHz. 
As occurs at cm-wavelengths, the quoted standard errors of the relative position are dominated by the 
uncertainties of the fixed parameters of the astrometric model (entries 3 to 10 of Table 1), 
and, in particular, by the uncertainties of the coordinates of the reference source, 
as expected 
for objects with a large angular separation (notice that this error is not frequency 
dependent). The comparison and interpretation of the relative 
position estimate at 43\,GHz with previously reported
estimates at lower frequencies will be postponed to a later 
publication where the comparison will be made in great detail.

\noindent
The postfit residuals of the differenced phase delays corresponding 
to all baselines included in our analysis are shown in Fig. 2. Note the scale 
of the plots, $\pm$23\,ps, corresponding to $\pm$1 phase cycle. The average 
root-mean-square (rms) of the postfit residuals is 2\,ps, less than one tenth of 
the phase cycle at 43\,GHz. At this level of 
precision, the absence of systematic effects validates both the
astrometric model, based on IERS 
standards, and the propagation medium procedures for mm-wavelength VLBI astrometry 
(at least for cycle times, source 
separations, weather conditions, and antenna elevations similar to those presented 
in this paper). To calibrate 
the quality of our procedure, we compared the residual 
of the differenced phase delay with similar 
residuals corresponding to observations at 8.4\,GHz and 5\,GHz
made in the past (G95, G98). The rms of 
the residuals are about 15, 9, and 2\,ps for the data sets at 5, 8.4, and 43\,GHz, 
respectively, which expressed in equivalent-phase yield 
postfit residuals of roughly 30 degrees at each of the three frequencies. 
This similarity of the rms expressed in phase at all the observed frequencies 
demonstrates not only that the phase connection process is feasible at 43\,GHz, 
but which is also of no less quality than at lower frequencies.\\
Likely, the most important contributors to the scatter of the phases at 43\,GHz 
are the unmodeled variations of the refractivity of the neutral atmosphere.
From the average rms of 2\,ps of the phase residuals of Fig. 2, and assuming 
uncorrelated contributions from the antennas forming each interferometric 
pair, the average uncertainty for the single-site phase delay 
is $\sqrt{2}$\,ps. This uncertainty is in good agreement with 
the predictions of water vapor fluctuations on time scales of 
100 seconds ($\sim$2\,ps) based on refractivity patterns 
described by Kolmogorov turbulences (Treuhaft \& Lanyi 1987).\\ 
The importance of our result translates to VLBI phase-referencing mapping. 
This technique (see e.g. Lestrade \et 1990) relies completely on the behavior 
of the phase of the reference source (usually a strong radio emitter) to 
predict the phase of the target source (usually a weak radio emitter). 
Beasley \& Conway (1995) provide useful expressions for the maximum cycle time 
for phase-referencing with the VLBA.  Under good weather conditions, average 
antenna elevation of 40\degr, and with the requirement that the rms phase between scans is
$<$90\degr, the maximum cycle time at 43\,GHz is $\sim$100s. This estimate 
should be shortened if atmospheric spatial variations from different lines of 
sight are considered. Actually, the facts are more favorable. 
For sources separated 5\degr\, on the sky, high antenna elevations, and good 
weather conditions, our results show that 
(i) the rms of the phases is below 90\degr\, throughout the experiment and 
does not seem to be substantially 
dependent of the cycle times used during our observation (100-300s); and 
(ii) the expected average uncertainty in interpolating the phases of one source to 
the epoch of the other is $\sim$30\degr\,in the differenced phase. This value 
is not larger than usual 
phase errors in phase-reference mapping at cm-wavelengths (e.g. Lestrade \et 1990).
Therefore, with the proper cycle times and nearby calibrator sources, diffraction 
limited VLBI phase-reference images at 43\,GHz should be possible.

\begin{figure}[t]
\vspace*{9cm}
\includegraphics{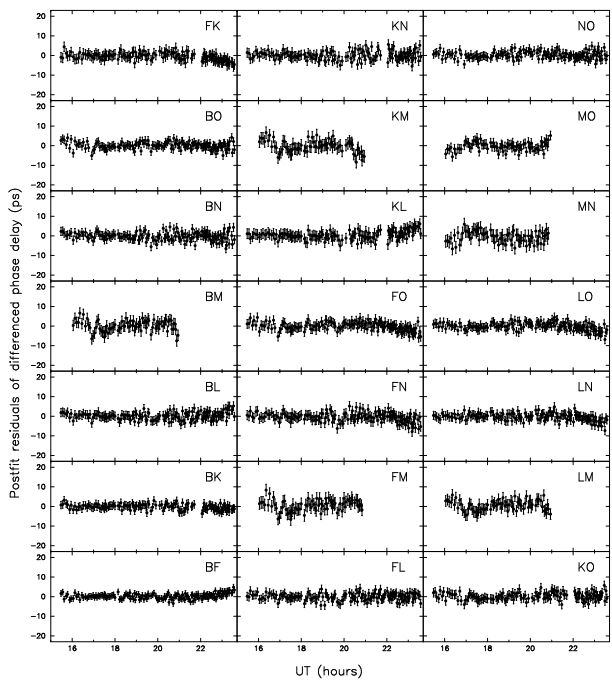}
\caption[]{Postfit residuals of the difference phase delays at 43\,GHz for all baselines.
Full vertical scale is $\pm$one phase cycle ($\pm$2$\pi$ equivalent), i.e., 
$\pm$23\,ps. The average rms is 2 ps, less than one tenth of a phase cycle at 43\,GHz. 
The symbols correspond to the following VLBA antennas: B, Brewster; 
F, Fort Davis; K, Kitt Peak; L, Los Alamos; M, Mauna Kea; N, North Liberty; 
O, Owens Valley} 
\end{figure}

\noindent
Our observations have shown that VLBI differential astrometry at 
43\,GHz provides high-precision relative positions. At this frequency, 
the astrometric precision is nearly equivalent to the resolution of the 
maps, and the reference point selected in the 
source structure might be associated with 
the core. This makes 43\,GHz differential astrometry an ideal technique 
to trace unambiguously the kinematics of the inner regions of the 
extragalactic radiosources. 

\acknowledgements{We thank Patrick Charlot for a constructive 
refereeing of the paper and Walter Alef for his valuable comments. 
We thank Jon Romney for his efforts during the 
correlation. This work has been supported by
the Spanish DGICYT grant PB96-0782. The National Radio Astronomy Observatory is 
operated by Associated Universities Inc., under a cooperative agreement with the 
National Science Fundation.}

\end{document}